# Computational Natural Philosophy: A Thread from Presocratics through Turing to ChatGPT

**Gordana Dodig-Crnkovic** [1,2,*]

[1] Department of Computer Science and Engineering, Chalmers University of Technology, 412 96 Gothenburg, Sweden

[2] School of Innovation, Design and Engineering, Computer Science Laboratory, Mälardalen University, Sweden

**Abstract:** This article examines the evolution of computational natural philosophy, tracing its origins from the mathematical foundations of ancient natural philosophy, through Leibniz's concept of a "Calculus Ratiocinator," to Turing's fundamental contributions in computational models of learning and the Turing Test for artificial intelligence. The discussion extends to the contemporary emergence of ChatGPT. Modern computational natural philosophy conceptualizes the universe in terms of information and computation, establishing a framework for the study of cognition and intelligence. Despite some critiques, this computational perspective has significantly influenced our understanding of the natural world, leading to the development of AI systems like ChatGPT based on deep neural networks. Advancements in this domain have been facilitated by interdisciplinary research, integrating knowledge from multiple fields to simulate complex systems. Large Language Models (LLMs), such as ChatGPT, represent this approach's capabilities, utilizing reinforcement learning with human feedback (RLHF). Current research initiatives aim to integrate neural networks with symbolic computing, introducing a new generation of hybrid computational models. While there remain gaps in AI's replication of human cognitive processes, the achievements of advanced LLMs, like GPT4, support the computational philosophy of nature – where all nature, including the human mind, can be described, on some level of description, as a result of natural computational processes.

## 1. Historical Roots of Computational Natural Philosophy

The predecessors of modern computational natural philosophy can be traced back to the ancient Greeks, pre-Socratics Thales, Pythagoras, Plato, and Aristotle who were pioneers in the development of Mathematics and Physics, as a means for discovering the laws of nature and the world. The pre-Socratic philosophers perceived the world as a structured system, a cosmos, that could be understood through logical thinking. They concentrated on how everything fit together logically, dismissing supernatural elements. Instead, they preferred to explore the natural laws that operate in the world and within human society. These philosophers were familiar with the computational knowledge of ancient Egypt and Babylon civilizations. Many of the results of this period are the basis of natural sciences like Euclidean geometry, Pythagoras theorem or Aristotle's system of logical deduction that became the basis for many later scientific and philosophical developments (Copeland and Proudfoot 1996).

Over the years, natural philosophy was only viewed as one aspect of philosophy, rather than a unique area of study. However, in the 16th century, René Descartes emerged as a key figure in the evolution of computational natural philosophy. He firmly believed that the universe could be understood through mathematics and logical reasoning. He developed a system of analytic geometry that allowed geometric problems to be solved algebraically, and his approach to philosophy had a major influence on the development of modern science.

In the 17th century, Blaise Pascal and Gottfried Wilhelm Leibniz were major predecessors of modern computational natural philosophy. Pascal constructed the first mechanical calculator, while Leibniz improved it and among others invented the binary system of counting, which forms the basis of modern computing. Galileo Galilei has been called the father of modern science. His greatest contribution was the "mathematization" of science. In 1623 Galileo wrote that Nature is a book written in "the language of mathematics". One of the greatest natural philosophers of his time, Isaac Newton published "The Mathematical Principles of Natural Philosophy", which exposed Natural philosophy as the "great mother of the sciences," including physical, chemical, and biological sciences.

Continuing in the improvement of computational devices, in the 19th century, Charles Babbage and Ada Lovelace developed the concept of the "analytical engine," a machine that could perform mathematical calculations through a series of punched cards. This idea laid the foundation for modern computing which is the basis of modern sciences and contemporary philosophy of nature.

The history of computational natural philosophy spans many centuries and involves generations of philosophers and mathematicians. Natural philosophy served as the foundation for the development of natural sciences. The transition from natural philosophy to natural science took place gradually over several centuries, with a significant shift occurring during the Scientific Revolution from the 16th to the 18th century. This period marked a change in the approach to studying the natural world, shifting the focus from answering the question "why?" to understanding "how?". Scientists began emphasizing observation, measurement, and systematic experimentation as methods to comprehend the workings of nature. This transformation led to the establishment of disciplines such as physics, chemistry, biology, and other natural sciences.

While natural philosophy, characterized by speculative reasoning and philosophical contemplation, continued to exist as a broader field encompassing philosophical inquiries into the fundamental principles of nature, the natural sciences experienced significant growth and specialization. The natural sciences thrived through rigorous study based on empirical evidence, scientific methodology, and practical applications.

Within the context of natural philosophy as a precursor to the natural sciences, computation played a dual role. It served as a tool for calculating and predicting the behaviors of nature, as well as providing explanations for the underlying mechanisms behind those behaviors. The fundamental notion that the universe, including humans, can be understood through mathematics, computation, and logical reasoning has remained a constant throughout the history of natural philosophy.

**2. Leibniz as a Forerunner of ChatGPT**

Gottfried Wilhelm Leibniz is considered by many to be one of the most important founders of computational natural philosophy through his contributions as a philosopher, mathematician, and polymath in the development of calculus, logic, and the philosophy of mind.

Leibniz envisaged the idea of a comprehensive project for human knowledge, Mathesis Universalis (universal science/encyclopedia of all human knowledge), which built upon a universal computational language Characteristica Universalis that would allow people to communicate with each other across linguistic and cultural boundaries. He believed that such a language should be based on a calculus of reasoning "Calculus Ratiocinator, a formal system of rules for reasoning about logical propositions. He proposed that such a system could be used to solve all problems that could be solved by means of logical deduction.

Leibniz's ideas about computation and natural philosophy were very influential in the development of modern computer science and artificial intelligence. His work on logic and formal systems provided a foundation for the development of programming languages and computer algorithms, while his ideas about the nature of the mind continue to inspire researchers in the field of cognitive science and artificial intelligence.

The present-day success of ChatGPT (GPT = Generative Pre-trained Transformer) may be seen as the beginning of the realization of Leibniz's dream about encyclopedia of all human knowledge built and ran by computational means corresponding to "calculus of reasoning" operating over "all information and knowledge of humanity" contained in LLMs.

### 3. Turing on the Computing Nature

Alan Turing is recognized for his contributions to the theory of computation (Turing Machine model) (Turing 1950), AI (as the first researcher in the field) (Christof Teuscher 2012), and connectionist models of learning in the form of "unorganized machines" (neural networks) (Boccato et al. 2011; Christof Teuscher 2012; Turing 1948). He made fundamental insights into the role of learning for intelligence, testing the ability of a machine to imitate human dialogue (Turing Test/Imitation game), and computational models of pattern formation (Turing 1952). However, his biographer Andrew Hodges sees Turing primarily as a natural philosopher: *"His thoughts and life were a generation ahead of his time, and the aspects of his ideas that transcended the 1940s' boundaries are best captured using the old-fashioned term: natural philosophy."* (Hodges 1997)

It is worth noticing that Turing's natural philosophy extends beyond Galileo's belief that "the book of nature is written in the language of mathematics" (The Assayer 1623) (Galilei 1623). Computing in a sense exceeds mathematics as computers not only represent numbers and relationships but also can generate real-time behaviors. Turing investigated numerous natural phenomena and suggested their computational models. His cutting-edge work on the relationship between computation and intelligence, as well as his research on morphogenesis as the formation of natural patterns, exemplify the natural philosopher's approach. In his pioneering 1952 paper on morphogenesis, Turing introduced a chemical computational model as the foundation of biological patterns that is still referred to in biology, (Galilei 1623).

Turing did not initially suggest that the physical system generating patterns performs computation through morphogenesis. However, from the standpoint of modern natural computationalism and specifically within the context of info-computation (Dodig-Crnkovic 2012a, 2012b, 2013a, 2014a, 2014b, 2017a; Dodig-Crnkovic and Giovagnoli 2012) it can be argued that morphogenesis represents a process of morphological computing. Though not computational in the conventional sense, *physical processes embody natural (unconventional), physical, intrinsic, morphological computation.* Morphology refers to the study of the structure, form, and arrangement of parts within a whole.

A critical component in modeling this process of intrinsic computing of nature is the interaction between informational structures and computational processes - the self-structuring of embodied information. The computation process in nature embodies physical laws on informational structures, and through this process, structures dynamically change their forms, as described in (Dodig-Crnkovic 2012a; Dodig-Crnkovic and Giovagnoli 2012). At some level of abstraction, all computation can be described as morphological computation - a process that changes or generates forms (Dodig-Crnkovic 2012b, 2013a, 2014a, 2014b, 2017a, 2023).

### 4. The Conceptual Evolution of Modern Computational Natural Philosophy, from Leibniz to Turing to ChatGPT

The history of modern computational natural philosophy is marked by milestones that demonstrate the continual evolution of the field, from its beginnings in Leibniz's idea of universal science/universal encyclopedia of knowledge combined with universal language and calculus of reasoning to Turing's ideas about morphogenesis and the nature of learning, to the present-day accomplishments of AI models such as ChatGPT. This article explores the development of computational natural philosophy, highlighting breakthroughs as well as key criticisms that have shaped the discipline.

Alan Turing's development of the Turing Machine, a theoretical model of computation as symbol manipulation, was followed by his concept of an "unorganized machine" (neural network, continuous model) and his proposition of the Turing Test/Imitation Game, designed to assess machine intelligence.

The concept of an "unorganized machine", suggested in the 1948 report, hypothesized that the infant human cortex operates as such a machine. These unorganized machines were randomly constructed but capable of *being trained to perform specific tasks* and were early examples of randomly connected binary neural networks. Turing proposed two types of unorganized machines: A-type machines, which were essentially randomly connected networks of NAND logic gates, and B-type machines, which were created by replacing inter-node connections with a structure called a connection modifier to allow for "appropriate interference, mimicking education" and to organize the network's behavior to perform useful work. Turing believed that the behavior of B-type machines could be complex, especially when the number of nodes in the network was large. Observe the idea of *training the unorganized machine*.

A major development in computational natural philosophy was the emergence of connectionism, which modelled intelligent behavior using artificial neural networks inspired by the human brain. The backpropagation algorithm, a key innovation in this area, enabled neural networks to learn from data and make predictions, enabling more complex AI models and deep learning. The advent of deep learning marked a significant turning point. With the introduction of powerful algorithms and the availability of vast amounts of data, AI models began to display extraordinary capabilities in tasks such as image recognition, natural language processing, and game playing. Notable examples include DeepMind's AlphaGo and OpenAI's GPT series.

The development of ChatGPT and GPT4 represents a major milestone in the evolution of the computational basis of contemporary computational natural philosophy. As an advanced large language model, ChatGPT shows human-like abilities in generating coherent and contextually relevant responses. This breakthrough has significant implications for the understanding of the mind, offering new opportunities for interpretation and simulating intelligent behavior. It has still numerous weaknesses like lack of robustness or tendency to fabulate ("hallucinate") but development goes very fast and present state of the art is certainly not the last word in the LLMs development.

**5. Computational Natural Philosophy and Naturalist Computationalism**

In the tradition of classical natural philosophy, modern computational natural philosophy relies on computation as both a methodological and conceptual tool, with the latter one becoming more and more prominent. In 1967, Konrad Zuse first proposed the idea that the entire universe might be functioning as a computational system at a fundamental level, using cellular automata as a model. He called this concept Computing Space ("Rechnender Raum") (Zuse 1970), which made him the pioneer in the theory field known as pan-computationalism/natural computationalism/computing nature. Steven Wolfram (Wolfram 2002, 2020) supports a pan-computationalist perspective, advocating for a dynamic type of reductionism where the complexity of natural behaviors and structures arises from a limited set of basic structures and processes. As a result, natural phenomena are products of computational processes. With similar premises, Edward Fredkin (E. Fredkin 1990; Edward Fredkin 2003) put forth his Digital philosophy, suggesting that cellular automata could give rise to particle physics.

Wolfram and Fredkin, following in Zuse's footsteps, considered the universe to be a fundamentally discrete system. However, the computing universe hypothesis (natural computationalism/computing nature) does not depend on the discreteness of the physical world. There are various types of computations in nature, including digital, analog, discrete, and continuous-state, see the taxonomy by Dodig-Crnkovic and Burgin (Burgin and Dodig-Crnkovic 2015). At the quantum-mechanical level, the universe carries out both continuous and discrete computations on dual wave-particle objects as argued by Lloyd (Lloyd 2006).

Even though the majority of computational theories assume discrete computing nature, the neural network models compute continuous-valued data. On the other hand, classical computation of the Turing Machine Model type, is based on (discrete) human reasoning, involving symbol manipulation, on which Turing built abstract thinking, knowledge composition, generalization, and efficient learning. Integrating both discrete and continuous representations emerges as crucial for developing systems with a general form of intelligence, as argued in (Cartuyvels et al. 2021).

Greg Chaitin has made significant contributions to the field of computing nature, beginning with information theory and computation as information processing. One of his key areas of focus has been information compression, which he famously described as "*compression is comprehension*" (G. Chaitin 2006)). This idea is particularly relevant to Language Models (LLMs) and underscores their fundamental importance. Chaitin's work has led him to the conclusion that the world can be understood and constructed through information and computation. He has also put forward the notion of *life as evolving software,* (G. Chaitin 2007, 2012; G. J. Chaitin 2018).

In (Zenil 2012), contemporary philosopher of nature Hector Zenil is exploring both the *nature of computation* and *nature as computation*.

## 6. Natural/ Physical/ Morphological/ Cognitive Computation: Sub-symbolic and Symbolic

The concept of natural computation as presented by (Dodig-Crnkovic 2013b; Rozenberg et al. 2012) addresses information processing (both discrete and continuous), as intrinsic characteristics of nature (Crutchfield et al. 2010). Models of natural computation/natural information processing differ from the Turing model of computation, that refers exclusively to symbol manipulation. From the point of view of the organization of computational processes, natural computation in general is different from von Neumann's (conventional) computation. Of course, all physical processes including those in technological devices perform natural computation. In the words of Greg Chaitin: "And how about the entire universe, can it be considered to be a computer? Yes, it certainly can, it is constantly computing its future state from its current state, it's constantly computing its own time-evolution! And as I believe Tom Toffoli pointed out, actual computers like your PC just hitch a ride on this universal computation!" (G. Chaitin 2007)

Natural computation models of biological organisms with their multi-level processes of computing and distributed information communication are capable of capturing the dynamic behavior of natural cognitive agents, including neuronal networks (Buzsáki 2009), for which the Turing Machine Model as a small subset is inadequate.

In the framework of info-computational nature, the fundamental mechanism is *morphological computation*, i.e. a process of transformations of the structure, form, and arrangement of parts within a whole. Those processes can be described as information self-organization (Hermann Haken 2006, 2008; Hermann Haken and Portugali 2017). For cognitive agents, morphological computing in living nature is a network of informational processes with cognition as layered morphological computation. Josh Bondgard and Michael Levin use the term "*Polycomputing*" (Bongard and Levin 2023). The author addressed this topic in (Dodig-Crnkovic 2012c, 2013b, 2017b, 2017c, 2020) as a computing on different levels of organization in nature.

In robotics, specific use of the term "morphological computation" has been adopted to denote decentralized embodied control of robots, where appropriate body morphology saves central information processing resources and enables learning through the self-structuring of information in a cognitive (robotic) agent, as described by Pfeifer et al. (Hauser et al. 2014; Pfeifer et al. 2006, 2007). This is a macroscopic view of morphological computation that does not concern lower levels of organization such as cellular, molecular, or quantum computation.

In a broader sense, morphological computation is a fundamental mechanism of all info-computational nature, which is the process of information self-organization as shown by Haken (H. Haken 1987; Hermann Haken 2006, 2008) and Haken and Portugali, (Hermann Haken et al. n.d.). Morphological computing in living nature as a subset of all physical computation is a network of morphological informational processes for cognitive agents, with cognition as layered morphological computation. This topic has been addressed in detail by Dodig-Crnkovic in several works (Burgin and Dodig-Crnkovic 2013; Dodig-Crnkovic 2014c, 2016; Dodig-Crnkovic and Stuart 2007).

Natural computation appears on all levels of organization in nature, from physical, chemical, and biological to cognitive. Of special interest for us are the levels of chemical and biological computation contributing to cognitive behavior such as presented by (Lones et al. 2013), showing the biochemical basis of connectionism. On the level of cells and tissues, there are numerous computational approaches such as (Lyon et al. 2021a; Turner et al. 2013; Tyrrell et al. 2016a). Proposals have been made for synthetic analog computation of living cells with artificial epigenetic networks (Daniel et al. 2013). Extensive literature exists on specific neuronal computation (Piccinini 2020; Piccinini and Shagrir 2014), as well as computational models of the brain, (Averbeck et al. 2006; Barron et al. 2020; Buzsáki and Draguhn 2004; Hertz et al. 2018; Laughlin and Sejnowski 2003; Sardi et al. 2017).

It is important to keep in mind the difference between new computational models of intrinsic information processes in nature (natural computing/morphological computing), and old computationalism based on the computer metaphor of the Turing Machine, performing symbol processing, which has been rightly criticized as an inadequate model of human cognition (Miłkowski 2018a; Scheutz 2002).

In humans, according to Daniel Kahneman, there are two basic cognitive systems, System 1 (reflexive, non-conscious, automatic, intuitive information processing, which is fast) and System 2 (reflective, conscious, reasoning and decision making, which is slow) (Kahneman 2011; Tjøstheim et al. 2020). Recognizing only symbolic information processing leaves the symbol grounding problem open. Sub-symbolic data/signal processing in deep learning provides mechanisms of symbol grounding.

Hybrid symbolic-dynamical (sub-symbolic) models (Bekkum et al. 2021; Larue et al. 2012) have been proposed, capable of modeling a combination of the two Kahneman's systems as a reactive-deliberative behavior. According to (Ehresmann 2014), the fast reflexive System 1 can be understood in terms of Rovelli's (Rovelli 2015) physical correlations (Shannon's relative information), and it can accommodate emotion as argued in (von Haugwitz et al. 2015). The slow System 2, because of synonymity in the symbol system, introduces an element of choice and indeterminism. The latter has been addressed in (Mikkilineni 2012), also tackling the topic of parallel concurrent computation typical of biological systems, for which the Turing Machine model is not sufficiently expressive.

**7. Contemporary Research in Computing Nature**

At present, computational natural philosophy is an active field based on research exploring the computational aspects of nature on its different levels of organization. The concept of natural computation, introduced by Rozenberg, Bäck, and Kok (Rozenberg et al. 2012), describes information processing in nature, encompassing both discrete and continuous processes (Dodig-Crnkovic and Giovagnoli 2013). Natural computation models differ from the classical Turing model of computation, which is solely based on symbol manipulation. According to (Burgin and Dodig-Crnkovic 2013, 2015), natural computation involves multi-level processes of computing and distributed information communication, capable of capturing the dynamic behavior of natural cognitive agents, including neuronal networks as described by Buzsáki (Buzsáki 2009).

Natural computation manifests itself at various levels of organization in nature, including the physical, chemical, biological, and cognitive domains. Of particular interest are the levels of chemical and biological computation that contribute to cognitive behavior, as exemplified by Lones et al. (Lones et al. 2013), who elucidated the biochemical basis of connectionism. At the cellular and tissue levels, numerous computational approaches have been proposed, such as those presented by (Tyrrell et al. 2016b), (Turner et al. 2013), (Lyon et al. 2021b), and (Levin et al. 2021). Additionally, Daniel et al. (Daniel et al. 2013) have put forth proposals for synthetic analog computation of living cells using artificial epigenetic networks. Extensive literature exists on specific neuronal computation, including the works of Piccinini and Shagrir (Piccinini 2020; Piccinini and Shagrir 2014), as well as computational models of the brain, as presented by Laughlin and Sejnowski (Laughlin and Sejnowski 2003), Buzsáki and Draguhn (Buzsáki and Draguhn 2004), Averbeck, Latham, and Pouget (Averbeck et al. 2006), Sardi et al. (Sardi et al. 2017), Hertz, Krogh, and Palmer (Hertz et al. 2018), and Barron et al. (Barron et al. 2020).

## 8. Criticisms of the Computational Nature Based on the "Triviality of Computation" Argument

In his work on Physical Computation Gualtiero Piccinini (Piccinini 2017, 2020; Piccinini and Anderson 2018), the most prominent critic of the idea of computational nature, presents several critical arguments against Pancomputationalism/Naturalist computationalism/Computing Nature. Unlimited Pancomputationalism, the most radical version of Pancomputationalism, according to Piccinini asserts that "every physical system performs every computation—or at least, every sufficiently complex system implements a large number of non-equivalent computations". In response, I argue these to be two substantially different claims. The first one, that every system executes every computation, has no support in physics and other natural sciences. Different sorts of systems perform different sorts of dynamical behaviors. In living systems we found "*Polycomputing*" (Bongard and Levin 2023). The second claim, that sufficiently complex systems implement a large number of different computations, is in accordance with natural sciences and fundamentally different from the claim that every system performs every computation which is false (Dodig-Crnkovic and Müller 2011a).

As for the sources of Pancomputationalism, Piccinini identifies the following: One source is "a matter of relatively free interpretation "which computation a system performs. This may well be true of human computational devices like fingers, pebbles, abacuses, and computers even though interpretations once chosen are kept constant (thus no longer free), in order to allow social communication of results.

Another source of pancomputationalism is the causal structure of the physical world. That claim goes one step further than the first one, actually searching for the basis of "free interpretation". We can freely choose systems used for calculation/computation, but the computational operations performed are predictable because of the laws of physics which guarantee that physical objects behave in the same way and according to physical laws so that we can predict and use their behavior for computation.

Info-computationalism is in the Piccinini scheme based on the third source:

"A third alleged source of pancomputationalism is that every physical state carries information, in combination with information-based semantics plus a liberal version of the semantic view of computation. *According to the semantic view of computation, computation is the manipulation of representations*. According to information-based semantics, *a representation is anything that carries information*. Assuming that every physical state carries information, it follows that every physical system performs the computations constituted by the manipulation of its information-carrying states (cf. Shagrir 2006). "

In Piccinini's view (Piccinini 2017)(Piccinini and Anderson 2018), both information-based semantics and the assumption that every physical state carries information are questionable. On the contrary, in the framework of computing nature/natural computationalism/pan-computationalism (Dodig-Crnkovic 2017b), for a cognizing agent, natural or artificial, the physical world presents an informational structure with computational dynamics.

Notwithstanding Piccinini's skepticism, there are well-established theories in computer science that do exactly the job of connecting computational processes and informational structures as suggested by info-computationalism, (Dodig-Crnkovic 2017b).

In his later work, Piccinini made a substantial move in the direction of natural computationalism by advocating, what he calls "*the modest view of the physical Church-Turing thesis*" (Piccinini 2017). In yet further developments of this approach, Piccinini details his neuro-centric position (Piccinini 2020) (Piccinini 2022). His claim now is that not all of the physical computation is of the Turing machine type. This view agrees with our best knowledge about natural computation today and it also brings us closer back to Turing's visionary work of computing nature.

Vincent Müller expressed his criticism of computing nature in a dialogue article with the author (Dodig-Crnkovic and Müller 2011b), arguing that if computation is everywhere and applies to everything, it does not explain anything. The response of the author was that computation appears in different forms at different levels of organization of matter, in a similar way as matter presents a universal principle but it is diversified at different levels of organization and therefore presents a useful theoretical tool for understanding the nature.

A similar example of a critical view toward computing nature (computing universe or pan-computationalism) is formulated by Marcin Miłkowski who admits that computation plays an important role but nonetheless cannot be a universal principle (Miłkowski 2007, 2013a, 2013b, 2018b, 2018c). My argument again is that info-computation is universal in the sense of matter-energy as basic principles that apply generally in nature.

**9. Penrose´s Criticism of Classical Computationalism and the Absence of Consciousness in ChatGPT**

Roger Penrose has expressed criticism of computationalism in two of his books and proposed his alternative perspectives on the nature of consciousness and human cognition. The first one, "*The Emperor's New Mind: Concerning Computers, Minds, and the Laws of Physics*" (Penrose 1989) explores the limitations of artificial intelligence and computational models in explaining human consciousness and understanding. He refers to Gödel's incompleteness theorems, the nature of consciousness, the role of quantum mechanics in the brain, and the limitations of algorithmic reasoning. In the second book "*Shadows of the Mind: A Search for the Missing Science of Consciousness*" (Penrose 1994) Penrose presents a continuation of the exploration of consciousness and its relationship to computation. He expands upon his previous arguments and addresses the criticisms and responses received after the publication of "The Emperor's New Mind." Penrose here suggests quantum processes underlying consciousness.

In the more recent work, from 2012, "Foreword to A Computable Universe Understanding Computation & Exploring Nature As Computation" (Zenil 2012) which is Penrose's latest text on computationalism, he explicitly admits that he changed his position in question several times. Penrose in the Foreword writes about different versions of computationalism and different possible interpretations. His critical position, explicitly stated in the Foreword is based on the Turing Machine Model of computation, thus under the assumption: Computation = Turing Machine = Algorithm.

When Penrose writes that consciousness is uncomputable he argues it is not algorithmic in the sense of the Turing Machine. That is a well-established position even within modern computational approaches. Instead, distributed asynchronous and concurrent models of computation are necessary.

As pointed out before, it is very important to distinguish between new computational models of intrinsic information processes in nature, such as natural computing/ morphological computing, and old computationalism based on the Turing Machine Model of computation, which performs symbol manipulation. This model has been criticized as inadequate for modeling human cognition (Miłkowski (Miłkowski 2007)(Miłkowski 2018b) Scheutz (Scheutz 2002), as well as irrelevant for AI, Sloman (Sloman 2002).

**10. Large Language Computational Models Passing the Turing Test**

Turing's proposal of the possibility of constructing artificially intelligent agents based on computations performed by electronic machines was met with skepticism. He also proposed the Imitation Game/ Turing Test as assessment of "machine intelligence" in a sense of ability of a machine to simulate/imitate human (written) language. Today's speech generation technology makes it possible to remove requirement of written communication, as the transformation to spoken language is not a problem for modern technology.

Human intelligence is even to this day by many considered to be impossible to implement in machines, assumed as being substantially different in nature, and supposed to be "non-material" or at least "non-mechanistic".

Recent technological advances in the development of Large Language Models (LLMs), ChatGPT, and other GPT (Generative Pretrained Transformer) programs capable of believable "intelligent" dialogues with humans finally present justification for Turing's belief in the possibility of the realization of intelligence in machines, i.e., Artificial Intelligence (AI). Even though GPTs are far from perfect, the goal of making them even more "human-like" does not seem like an impossible goal at all.

Based on simple principles, known within the AI for several decades, from the beginning of deep neural networks, recent developments differ only in the following. They are applying much bigger neural networks, trained on a vast amount of data, with thousands of fast GPUs in large computer clusters, that run for several weeks, with elaborate infrastructure, optimization of neural networks, and architectures like Transformers. The final important step was adding *reinforcement learning with human feedback* (RLHF). In this process, AI human trainers provided a reward model in which AI responses were ranked by humans, so the AI learned which was the best response. The method is simple and effective: predicting the next word based on huge, compressed data that contain human knowledge on a topic collected from vast available Internet sources. Up to now, large networks with a huge number of parameters have been used in LLMs, building large data sets and designing algorithms to pass Turing Test.

No symbolic computing has been involved yet, just deep neural networks (DNNs). Researchers like Yoshua Bengio (Russin et al. 2020) believe in the power of *hybrid computing models* meaning involving both DNNs and symbolic computing. This necessity of two complementary systems is motivated by Kahneman (Kahneman 2011) and Tjøstheim et al. (Stephens and Tjøstheim 2020) two basic cognitive systems: fast, reflexive, unconscious, automatic, System 1, and slow, reflective, conscious, reasoning, and decision-making processes. System 2 . The LLMs correspond to System 1. Current developments in AI are continuing towards the goals of modeling System 2 symbolic reasoning (Russin et al. 2020). Even though Bengio's and Kahneman's interpretations of System 1 and 2 are not identical, as evident from the discussion between Lecun, Hinton, Bengio and Kahneman, ("AAAI-20 Fireside Chat with Daniel Kahneman" 2020), the details are not essential for our argument. Even OpenAI's Sam Altman recently claimed that the age of giant AI models is over and that new development strategies will be needed in the future, ("AAAI-20 Fireside Chat with Daniel Kahneman" 2020).

Recent success of LLMs came as a surprise to many people who expected intelligent responses in natural language to be something more complex and difficult or even impossible to model. As Stephen Wolfram says:

"Perhaps, one might have imagined, there's something more to brains than their networks of neurons—like some new layer of undiscovered physics. But now with ChatGPT we've got an important new piece of information: we know that a pure, artificial neural network with about as many connections as brains have neurons is capable of doing a surprisingly good job of generating human language." (Wolfram 2023)

Now it seems to (almost) suffice with computation over vast amounts of data/information/knowledge produced by humans and further processed by machines. GPT-3.5 was trained on Wikipedia, social media, news, books, and other documents across different topics published before the end of 2021, with the use of data compression and additional RLHF training, ("AAAI-20 Fireside Chat with Daniel Kahneman" 2020). Currently, extensions are made to add web search to the prompts and make on-line data available in real time.

It is important to emphasize that Turing Test which GPT programs can pass *is not a test of consciousness* but shows the ability of a machine to produce believable human-like dialogue, which says nothing about consciousness. Being on the level of Kahneman's System 1 those processes are unconscious. Consciousness appears first on the level of System 2. Those are the views of modern computationalism as well (Piccinini 2020; Piccinini and Shagrir 2014) (Lyon et al. 2021b) (Levin et al. 2021).

The question of generating new knowledge not from a huge corpus of existing human knowledge but *de novo* is a different issue. *LLMs produce new knowledge from the existing one*. That is why GPT programs are trained on huge amounts of humanly produced text which required consciousness and physical presence in the world while it was being created. The situation is similar to the Synthetic Biology which can produce a living cell by combining parts of disassembled cells. It is still not capable of producing a living cell *de novo* from a container with organic molecules that living cell is composed of.

**11. Short Summary of Modern Computational Natural Philosophy**

Computational natural philosophy refers to the use of computational methods to study and understand natural phenomena, such as the behavior of matter, energy, and living organisms. This interdisciplinary field combines knowledge from computer science, mathematics, physics, biology, cognitive, behavioral and other natural sciences to model and simulate complex systems.

The roots of modern computational natural philosophy can be traced back to Leibniz, whose idea of universal human knowledge with a universal language and rational calculus inspired the development of computing models and techniques. Scientists began utilizing mechanical computers to simulate physical processes and engage in computational thinking about nature. Today, the field relies on various scientific disciplines to gain a comprehensive understanding of nature. From modeling the behavior of particles in fluids to simulating the dynamics of ecosystems and societies, computational natural philosophy draws from a broad range of scientific endeavors.

One of the advantages of computational natural philosophy is its ability to study systems that are too complex or challenging to observe directly. This includes molecules, atoms, subatomic particles, and distant astronomical objects. Additionally, it allows scientists to test hypotheses and make predictions about natural phenomena without the need for costly or time-consuming experiments. By conducting simulations and generative modeling of natural systems, researchers can anticipate how these systems might behave under different conditions.

Computational natural philosophy holds great potential for providing new insights into the workings of the natural world and addressing pressing societal challenges like climate change and disease outbreaks. Its overarching advantage lies in its ability to facilitate the integration of specialized scientific knowledge about nature into a systemic and ecological framework, thereby enabling a holistic understanding from diverse perspectives.

**12. Conclusion: Computational Natural Philosophy, from Leibniz to Turing to GPTs and Beyond**

This perspective article discusses the connection between modern natural philosophy and computation, from Leibniz with Calculus Ratiocinator via Turing's computational models of learning to the development of GPTs.

Computational natural philosophy has its deep ancient roots in the works of pre-Socratic Greek natural philosophers Thales, Pythagoras, Plato, and Aristotle, the tradition continued by Descartes, Pascal, Leibniz, Galileo, Newton, Babbage, Lovelace, and others. Turing contributions to natural philosophy range from models of computation, neural networks and AI to his research on learning and morphogenesis. Turing's work laid the foundation for modern natural computationalism, which investigates morphological computation as a process that generates form.

The consistent theme throughout computational natural philosophy is the understanding of the universe through mathematics/computation and logical reasoning searching for general laws and regularities. Leibniz's work on universal encyclopedia of all human knowledge encoded in universal language and traversed with the help of calculus of reasoning, influenced modern computer science and AI. Turing's work laid the foundation for computational approach to knowledge generation/learning with the invention of the Turing Machine (symbol processing), followed by the concept of an "unorganized machine," (neural network model) and the Turing Test of artificial intelligence. ChatGPT represents the beginning of the realization of Leibniz's dream of computational universal knowledge.

The field of contemporary computational natural philosophy involves naturalist computationalism, which suggests that the universe can be modelled as a computational system at a fundamental level, based on natural (physical) computation, which involves information processing at various levels of organization in nature.

Key criticisms of computational natural philosophy from the point of view of computational models are that the notion of computation as a fundamental property of the physical universe is too general and thus trivial. They include Gualtiero Piccinini's opposition to pan-computationalism, Vincent Müller's and Marcin Miłkowski's argument that computation's ubiquity diminishes its explanatory power.

Another line of critical thought, focused on computational theories of mind, is Roger Penrose's criticism which presupposes that knowledge generation requires consciousness of knowledge-generating agents. We concur and point out that programs like GPT do not require consciousness in order to generate data, information, and knowledge based on huge corpora of existing human knowledge/information. But for the *generation de novo, of knowledge about the physical world, by autonomous AI agents, embodiment and consciousness will be necessary. More powerful computational models are needed than those implemented in ChatGPT*. Penrose also argues that the Turing model of computation is not adequate for explaining the human mind. With that position also the leading representatives of computationalism agree.

Despite these criticisms, computational natural philosophy has made significant contributions to understanding the complexities of the natural world and the development of AI systems that connect discrete and continuous representations.

The success of LLMs in producing intelligent responses in natural language was surprising to many who expected that a more complex approach would be needed for passing Turing Test.

Deep learning computation in LLMs corresponds to the fast, intuitive System 1 thinking, described by Daniel Kahneman. Researchers, such as Yoshua Bengio, are working towards modeling System 2 symbolic reasoning by exploring hybrid computing models that combine neural networks and symbolic computing which would add an important complementary cognitive layer to LLMs.

In sum, the success of contemporary computational natural philosophy is based on the use of advanced computational thinking and techniques to comprehend and investigate natural phenomena, including the behavior of matter, energy, and living organisms with humans and human societies. This interdisciplinary research field merges knowledge, insights, and techniques from computer science, mathematics, physics, biology, and other natural sciences with humanities, and other domains to create models and simulations of complex systems in their broad context and in relation to each other.

Recent advances in LLMs keep a promise of further fundamental developments. Among them is the proposal of the necessity, in the next step to move from passive AI to ones active in the physical world – embodied, embedded, and enacted, as Giovani Pezzulo et al. suggest in (Pezzulo et al. 2023). Yoshua Bengio was on the same track when suggesting a move towards the System 2 cognitive functions in AI that would among others include causality, as argued in "Towards Neural Nets for Conscious Processing and Causal Reasoning" (Bengio 2022). Similar discussion can be found in "Generalist AI Beyond Deep Learning", by Christoph von der Malsburg, Michael Levin, Joshua Bach (Malsburg et al. 2023).

Starting from the early stages of computing, modern computational natural philosophy now spans multiple domains. It holds significant potential for delivering new insights into the natural world and helping understand major challenges such as climate change and disease outbreaks from various viewpoints. It offers a cohesive platform for today's sciences, which often work in isolated silos, in a world that increasingly requires an interdisciplinary and comprehensive understanding of nature and society.

The article suggests that natural computationalism and the idea of computing nature present a solid foundation for further advances in knowledge production and a deeper understanding of nature, in terms of both computational tools and models. This framework provides valuable insights into complex natural and social phenomena and the potential evolution of more sophisticated AI systems, ultimately improving our understanding of the world and our interactions with it.

In summary, this article discusses the possibility that learning, and knowledge generation can be done computationally, as Leibniz originally believed. The success of ChatGPT and other large language models can be seen as proof of concept for Leibniz's ideas. The computational view of knowledge, from Leibniz via Turing to ChatGPT, presupposes information transformations of existing human knowledge corresponding to Kahneman's System 1 and requires no consciousness. Future developments point towards implementing properties of System 2.

**Acknowledgment**

This paper is based on the research supported by the Swedish Research Council grant MORCOM@COGS.**References**

AAAI-20 Fireside Chat with Daniel Kahneman. (2020). https://vimeo.com/390814190

Averbeck, B. B., Latham, P. E., & Pouget, A. (2006). Neural correlations, population coding and computation. *Nature Reviews Neuroscience*. https://doi.org/10.1038/nrn1888

Barron, H. C., Reeve, H. M., Koolschijn, R. S., Perestenko, P. V., Shpektor, A., Nili, H., et al. (2020). Neuronal Computation Underlying Inferential Reasoning in Humans and Mice. *Cell*. https://doi.org/10.1016/j.cell.2020.08.035

Bekkum, M. van, Boer, M. de, Harmelen, F. van, Meyer-Vitali, A., & Teije, A. ten. (2021). Modular Design Patterns for Hybrid Learning and Reasoning Systems: a Taxonomy, Patterns and Use Cases. *arXiv:2102.11965v1 [cs.AI]*, (23 Feb).

Bengio, Y. (2022). Towards Neural Nets for Conscious Processing and Causal Reasoning. https://www.youtube.com/watch?v=psfh1fk2Qig

Boccato, L., Soares, E. S., Fernandes, M. M. L. P., Soriano, D. C., & Romis Attux. (2011). Unorganized Machines: From Turing's Ideas to Modern Connectionist Approaches. *International Journal of Natural Computing Research*, *2*(4), 1–16. https://doi.org/10.4018/jncr.2011100101

Bongard, J., & Levin, M. (2023). There's Plenty of Room Right Here: Biological Systems as Evolved, Overloaded, Multi-Scale Machines. *Biomimetics*, *8*(110). https://doi.org/10.3390/biomimetics8010110

Burgin, M., & Dodig-Crnkovic, G. (2013). The Nature of Computation and The Development of Computational Models. In *Computability in Europe 2013 (CiE 2013) The Nature Of Computation, Univ. of Milano-Bicocca, 1-5 July 2013*. Milano.

Burgin, M., & Dodig-Crnkovic, G. (2015). A Taxonomy of Computation and Information Architecture. In M. Galster (Ed.), *Proceedings of the 2015 European Conference on Software Architecture Workshops (ECSAW '15). Cavtat, Croatia, 7–11 September 2015;* New York, USA: ACM Press. https://doi.org/10.1145/2797433.2797440

Buzsáki, G. (2009). *Rhythms of the Brain*. *Rhythms of the Brain*. Oxford, UK: Oxford University Press. https://doi.org/10.1093/acprof:oso/9780195301069.001.0001

Buzsáki, G., & Draguhn, A. (2004). Neuronal olscillations in cortical networks. *Science*. https://doi.org/10.1126/science.1099745

Cartuyvels, R., Spinks, G., & Moens, M. F. (2021). Discrete and continuous representations and processing in deep learning: Looking forward. *AI Open*, *2*, 143–159. https://doi.org/10.1016/J.AIOPEN.2021.07.002

Chaitin, G. (2006). The limits of reason. *Scientific American*, *294*(3), 74–81.

Chaitin, G. (2007). Epistemology as Information Theory: From Leibniz to Ω. In G. Dodig Crnkovic (Ed.), *Computation, Information, Cognition – The Nexus and The Liminal* (pp. 2–17). Newcastle UK: Cambridge